\documentclass[aps,prc,preprint,showpacs,superscriptaddress,groupedaddress]{revtex4}
\usepackage{amsmath}
\usepackage{graphicx,epsfig}
\usepackage{hyperref} 
\newcommand{\beqy}{\begin{eqnarray}}
\newcommand{\eeqy}{\end{eqnarray}}
\begin{document}
%%%%%%%%%%%%%%%%%%%%%%%%%%%%%%%%%%%%%%%%%%%%%%%%%%%%%%%%%%%%%
%%%%%%%%%%%%%%%%%%%%%%%%%%%%%%%%%%%%%%%%%%%%%%%%%%%%%%%%%%%%%
%%%%%%%%%%%%%%%%%%%%%%%%%%%%%%%%%%%%%%%%%%%%%%%%%%%%%%%%%%%%%
%
\title{Low-energy collective excitations in the neutron star inner crust }
\author{N. Chamel}
\affiliation{Institut d'Astronomie et d'Astrophysique,
Universit\'e Libre de Bruxelles - CP226, 1050 Brussels,  Belgium}
\author{D. Page}  
\affiliation{Instituto de Astronom\'ia, Universidad Nacional Auton\'oma de M\'exico,  Mexico D.F. 04510, Mexico}
\author{S. Reddy}
\affiliation{Institute for Nuclear Theory, University of Washington, Seattle, Washington 98195, USA}

%%%%%%%%%%%%%%%%%%%%%%%%%%%%%%%%%%%%%%%%%%%%%%%%%%%%%%%%%%%%%
\begin{abstract}
We study the low-energy collective excitations in the inner crust of the neutron star, where 
a neutron superfluid coexists with a Coulomb lattice of nuclei. The dispersion relation of 
the modes is calculated systematically from a microscopic theory including neutron band structure 
effects. These effects are shown to lead to a strong mixing between the Bogoliubov-Anderson 
bosons of the neutron superfluid and the longitudinal crystal lattice phonons. 
In addition, the speed of the transverse shear mode is greatly reduced as a large fraction of superfluid 
neutrons are entrained by nuclei. Not only does the much smaller velocity of the transverse mode increase the 
specific heat of the inner crust, it also decreases its electron thermal conductivity. These results 
may impact our interpretation of the thermal relaxation in accreting neutron stars. Due to strong mixing, 
the mean free path of the superfluid mode is found to be greatly reduced. Our results for the collective 
mode dispersion relations and their damping may also have implications for neutron star seismology.   
\end{abstract}
%%%%%%%%%%%%%%%%%%%%%%%%%%%%%%%%%%%%%%%%%%%%%%%%%%%%%%%%%%%%%

\pacs{21.10.Dr, 21.60.Jz, 21.65+f, 26.60.+c}

\maketitle\thispagestyle{empty}

%%%%%%%%%%%%%%%%%%%%%%%%%%%%%%%%%%%%%%%%%%%%%%%%%%%%%%%%%%%%%
\section{Introduction} 
\label{intro}
%%%%%%%%%%%%%%%%%%%%%%%%%%%%%%%%%%%%%%%%%%%%%%%%%%%%%%%%%%%%%

The crust of a neutron star represents only about $10\%$ of the star's radius and $1\%$ of its mass but is 
expected to play a key role in various observed astrophysical phenomena such as pulsar glitches, quasiperiodic 
oscillations in soft-gamma ray repeaters (SGRs), and thermal relaxation in soft x-ray transients~\cite{lrr}. 
The outer crust is primarily composed of pressure ionized atoms arranged in a regular crystal lattice and embedded 
in a highly degenerate electron gas. 
With increasing density, electrons become relativistic and the rapid growth of their Fermi energy drives nuclei 
to become neutron-rich due to electron captures (see e.g. Ref.~\cite{pgc11}). Eventually, at a density 
$\sim 4 \times10^{11}$ g/cm$^{3}$, some neutrons drip out of nuclei (see e.g. Ref.~\cite{onsi08,pcgc12}). This defines the 
boundary between the outer crust and the inner crust. ``Dripped'' neutrons in the inner crust are expected to become superfluid 
below a critical temperature of the order of $\sim 10^{10}$ K (see e.g. Ref.~\cite{bal05}). Despite the absence of viscous drag, 
the neutron superfluid can still be coupled to the crust due to non-dissipative entrainment effects arising from elastic Bragg scattering of dripped neutrons by the crystal lattice~\cite{cch05,car06}. Recent calculations have shown that in 
some regions of the inner crust only a very small fraction of dripped neutrons participate in the superfluid 
dynamics~\cite{cha05,cha12}. Consequently, the vibrations of the crystal lattice are expected to be strongly coupled to 
the collective excitations of the neutron superfluid~\cite{and09,pet10,cir11}. Collective excitations are particularly 
important for understanding thermal and transport properties of accreting neutron stars with temperatures in the range 
$T=10^{7}-10^{9}$ K~\cite{pr12}. 

In this paper, we study low-energy collective modes with large wavelengths compared to the typical internuclei distance. 
The existence of two longitudinal modes in the inner crust and the role of entrainment in determining the dispersion 
relations were first studied in Ref.~\cite{eps88} using a hydrodynamic approach. In this pioneering study, long-range 
perturbations on the superfluid flow induced by the lattice of nuclei were neglected. Here we show that they play a crucial 
role. The low-energy constants that depend on the microscopic properties of the inner crust are calculated in a consistent 
approach that properly incorporates the long-range correlations leading to entrainment effects, first discussed in Ref.~\cite{cha05}. 
We find that entrainment of superfluid neutrons by crustal nuclei greatly reduces the velocity of the two transverse (shear) 
modes, and this in turn enhances their contribution to the low temperature specific heat of the inner crust. Entrainment effects 
also induce a strong mixing between the longitudinal lattice phonons and the Bogoliubov-Anderson (BA) bosons~\cite{bog58,and58} of 
the neutron superfluid, splitting these modes 
into a high velocity global sound mode and a low velocity mode characterized by a relative motion between the neutron 
superfluid and the electron-ion plasma. These results should be also relevant for studies of global neutron star seismic modes 
with frequencies in the range $20-1000$ Hz, which could be excited during violent events such as giant flares in SGRs and binary 
neutron-star mergers.

In the following section we define our notations and present order of magnitude estimates of the relevant length and momentum
scales.
In Sec.~\ref{dynamics} we describe the low-energy, long-wavelength, collective excitation modes and their velocities. The microscopic model of Ref.~\cite{cha12} is employed in \S~\protect\ref{crust} to obtain quantitive values for these velocities.
Damping of these modes is briefly considered in Sec.~\ref{dissipation}.
In the Sec.~\ref{implications} we study how entrainment affects the inner crust specific heat, its thermal conductivity, and
its thermal relaxation time scale. We finally conclude in Sec.~\ref{conclusions}.

%%%%%%%%%%%%%%%%%%%%%%%%%%%%%%%%%%%%%%%%%%%%%%%%%%%%%%%%%%%%%
\section{Basic notations and physical scales}
\label{notations}
%%%%%%%%%%%%%%%%%%%%%%%%%%%%%%%%%%%%%%%%%%%%%%%%%%%%%%%%%%%%%

In what follows, we will assume that the inner crust of a neutron star is a perfect crystal. 
Each crustal layer will consist of a body-centered cubic lattice containing only one type of nuclide and will be characterized by 
$Z$, the \emph{total} average number of protons in the Wigner-Seitz (W-S) cell of the crystal lattice (a truncated octahedron); 
$A_{\rm cell}$ the \emph{total} cell average number of nucleons; $A$, the cell average number of nucleons \emph{bound} 
inside nuclei; and $A^\star$, the cell average number of nucleons \emph{entrained} by the solid crust. As shown in 
Ref.~\cite{cha12}, $A^\star$ is generally much larger than $A$ and close to $A_{\rm cell}$ due to the Bragg scattering 
of unbound neutrons by the periodic potential of the crystal lattice, which manifests itself in neutron band structure effects. 

We will indicate by $n = n_p + n_n$ the total average baryon number density, which is the sum of the average proton density $n_p$ 
and average neutron density $n_n$. Neutrons entrained by nuclei are effectively bound. Their density will be noted as
$n_n^{\rm b}$. By analogy with conduction electrons in ordinary solids, neutrons that are not entrained will be referred to as 
\emph{conduction} neutrons and their density will be noted as $n_n^{\rm c}$. 
As shown in Ref.~\cite{cha12}, the density $n_n^c$ is generally much smaller 
than the density  $n_n^{\rm f}$ of ``free'' or ``dripped'' neutrons. Because of Galilean invariance, we have
\beqy
\label{1}
n_n = n_n^{\rm b} +  n_n^{\rm c}\, .
\eeqy
These densities are related to $Z$, $A^\star$, and $A_{\rm cell}$ by 
\beqy
\label{2}
n_n^{\rm b}  = \frac{A^\star-Z}{A_{\rm cell}} \, n \, ,
\eeqy
and 
\beqy
\label{3}
n_p  = \frac{Z}{A_{\rm cell}} \, n \, .
\eeqy
The ion number density $n_{\rm I}$ is determined by 
\beqy
\label{4}
n_{\rm I} = \frac{n}{A_{\rm cell}}\, .
\eeqy
As discussed in Refs.~\cite{and09,pet10,cir11}, the mass density associated with lattice vibrations is given by
\beqy
\label{5}
\rho_{\rm I} = m \,(n_p+n_n^{\rm b}) = A^\star\, m\, n_{\rm I} \, ,
\eeqy
where $m$ is the nucleon mass (neglecting the small difference between neutron and proton masses), whereas the 
total mass density (neglecting the electron contribution) is 
\beqy 
\label{6}
\rho = m \, n= A_{\rm cell} \, m \, n_{\rm I}\, .
\eeqy

The typical length scale associated with the solid crust is the ion-sphere radius defined by
\beqy
\label{7}
r_{\rm I} = \left( \frac{3}{4 \pi n_{\rm I}}\right)^{1/3} 
\approx 75 \left( \frac{A_{\rm cell}/1000}{\rho_{12}} \right)^{1/3} \; \text{fm}
\eeqy
where $\rho_{12} =\rho/(10^{12} \text{g cm}^{-3})$.
The characteristic angular frequency and wave number of lattice vibrations is the ion angular plasma frequency 
\beqy
\label{8}
\omega_p = \sqrt{\frac{4\pi (Ze)^2 n_{\rm I}}{A^\star m}} = \sqrt{\frac{4 \pi e^2 n^2_p}{ (n_p+n^{\rm b}_n)\, m}}
\eeqy
and the Debye wave number 
\beqy
\label{9}
q_D = (6 \pi^2 n_{\rm I})^{1/3} 
\approx \frac{2.4}{r_{\rm I}}
\approx 0.03 \left( \frac{\rho_{12}}{A_{\rm cell}/1000}\right)^{1/3} \; \text{fm}^{-1} 
\, ,
\eeqy
respectively. The ion plasma temperature is defined by $T_p=\hbar \omega_p/k_{\rm B}$ ($k_{\rm B}$ being the Boltzmann constant).

The ultrarelativistic electrons found in the inner crust of neutron stars with density $n_e = Z n_{\rm I}$, 
are almost uniformly distributed~\cite{maru05} and are characterized by their Fermi wave number
\beqy
\label{10}
k_{\rm Fe} = (3 \pi^2 n_e)^{1/3} = \left(\frac{Z}{2}\right)^{1/3} q_D 
\approx  \frac{7}{r_{\rm I}} \, \left(\frac{Z}{50}\right)^{1/3}
\approx 0.1 \left( \frac{\rho_{12} \; Z/50}{A_{\rm cell}/1000}\right)^{1/3} \; \text{fm}^{-1} 
\, .
\eeqy
Small deviations of the electron distribution from uniformity 
are characterized by the electron Thomas-Fermi screening wave number
\beqy
\label{11}
q_{\rm TFe} = \sqrt{4\pi e^2 \frac{\partial n_e}{\partial \mu_e}}
= \sqrt{\frac{4 \alpha}{\pi}}\, k_{\rm Fe} 
\approx 0.1 \, k_{\rm Fe}
\approx  \frac{0.7}{r_{\rm I}} \,  \left(\frac{Z}{50}\right)^{1/3}
\eeqy
where $\mu_e = \hbar c \, k_{\rm Fe}$ is the electron chemical potential and
$\alpha \equiv e^2/\hbar c \approx 1/137$ the fine structure constant.

%%%%%%%%%%%%%%%%%%%%%%%%%%%%%%%%%%%%%%%%%%%%%%%%%%%%%%%%%%%%%
\section{Low-energy dynamics of the neutron-star inner crust}
\label{dynamics}
%%%%%%%%%%%%%%%%%%%%%%%%%%%%%%%%%%%%%%%%%%%%%%%%%%%%%%%%%%%%%

The equations governing the low-energy dynamics of a nonrelativistic neutron superfluid immersed in an elastic crust have been 
derived in Refs.~\cite{car06b,pet10,cir11}. The corresponding normal modes of oscillation can be found by considering small perturbations 
of the densities and currents from their equilibrium values and solving the resulting linearized hydrodynamic equations. The first two 
of these equations arise from the conservation of neutron and proton numbers
\beqy\label{12}
\frac{\partial \delta n_n}{\partial t} +n_n^c\pmb{\nabla} \cdot \delta \pmb{v_n} +n_n^b\pmb{\nabla} \cdot \delta\pmb{v_p}=0\, ,
\eeqy
\beqy
\frac{\partial \delta n_p}{\partial t} + n_p \pmb{\nabla} \cdot \delta\pmb{v_p}=0\, ,
\eeqy
where $\delta n_n$ and $\delta n_p$ are the perturbed neutron and proton densities, respectively while 
$\delta\pmb{v_n}$ and $\delta\pmb{v_p}$ are the perturbed neutron and proton velocities, respectively. In the following
we will consider oscillations characterized by wave vectors $q\ll q_{\rm TFe}$ so the crustal matter remains electrically 
neutral locally and $n_p=n_e$. Treating the neutron-star crust as an isotropic solid and using $i$, $j$, $k$ for coordinate space indices, 
the momentum conservation can be expressed as
\beqy\label{13}
m n_n^c \frac{\partial \delta v_{ni}}{\partial t} + \rho_{\rm I}\frac{\partial \delta v_{pi}}{\partial t} 
+n_n\nabla_i \delta\mu_n+L\nabla_i \delta n_n -\tilde K\nabla_i u_{jj} 
-2S\nabla_j\left(u_{ij}-\delta_{ij}\frac{1}{3}u_{kk}\right)=0\, ,
\eeqy
where $u_{ij}$ is the strain tensor, $\delta\mu_n$ the perturbed neutron chemical potential, $S$ the shear modulus, and 
$\widetilde K$ the bulk modulus of the electron-ion system
\beqy\label{14}
\widetilde K = n_p^2\left( \frac{\partial \mu_p}{\partial n_p} + \frac{\partial \mu_e}{\partial n_e}\right)\, ,
\eeqy
and the coefficient $L$ given by 
\beqy\label{15}
L=n_p\frac{\partial \mu_n}{\partial n_p}\, ,
\eeqy
takes into account the coupling of the neutron superfluid to the strain field. 
The condition for neutron superfluidity is embedded in Josephson's equation
\beqy\label{16}
\frac{\partial \delta \pmb{v_n}}{\partial t} +\frac{1}{m}\pmb{\nabla}\delta\mu_n=0\, .
\eeqy

The normal modes 
have the form of plane waves that vary in space and time as $\exp[{\rm i}(\pmb{q}\cdot\pmb{r}-\omega t)]$, where $\pmb{q}$ is the wave 
vector and $\omega$ the angular frequency. In an isotropic medium, the normal modes may be separated into transverse and longitudinal ones. 
In the long wavelength limit $q\rightarrow0$, the normal modes all have a soundlike dispersion relation, with $\omega=v q$, $v$ being the 
mode speed. The speed of the two transverse lattice modes is given by~\cite{and09,pet10}
\beqy
\label{17}
v_t =\sqrt{\frac{S}{\rho_{\rm I}}}\, .
\eeqy
Due to interactions between neutron and proton densities and currents, the BA bosons of the neutron 
superfluid with velocity $v_\phi$ are mixed with the longitudinal lattice phonons with velocity $v_\ell$. Neglecting the 
coupling of the neutron superfluid to the strain field, the resulting dispersion relation is given by~\cite{pet10,cir11}
\beqy
\label{18}
(\omega^2-v_\phi^2~q^2)(\omega^2-v_\ell^2~q^2)=g^2_{\rm mix} ~\omega^2~q^2\, ,
\eeqy
where the strength of the mixing is characterized by the parameter
\beqy
\label{19}
g_{\rm mix} = v_\phi \sqrt{\frac{n_n^{\rm b}}{n_p+n_n^{\rm b}}~\frac{n_n^{\rm b}}{n_n^{\rm c}}} \,,
\eeqy
first introduced in Ref.~\cite{cir11}. 
The velocity of the BA mode is 
\beqy
\label{20}
v_\phi=\sqrt{\frac{n_n^{\rm c}}{m}\frac{\partial\mu_n}{\partial n_n}}\, ,
\eeqy
whereas the velocity of the longitudinal mode of the lattice is
\beqy
\label{21}
v_\ell=\sqrt{\frac{\widetilde K+4S/3}{\rho_{\rm I}}}\, .
\eeqy
In the neutron-star crust, the electron contribution to the bulk modulus 
dominates, and the ion contribution can be safely neglected (see, e.g., Sec. 7.1 of Ref.~\cite{lrr}). 
As a result, $v_\ell$ is approximately given by~\cite{ash76}
\beqy
\label{22}
v_\ell \approx  \frac{\omega_p}{q_{\rm TFe}}=\sqrt{\frac{n_p}{n_p+n^{\rm b}_n}~\frac{n_p}{m}~\frac{\partial \mu_e}{\partial n_e}} \, .
\eeqy
Solving Eq.~(\ref{13}) we find that the eigenmode velocities are given by 
\beqy
\label{23}
v_{\pm} = \frac{V}{\sqrt{2}}\sqrt{1\pm\sqrt{1-\frac{4 v_\ell^2 v_\phi^2}{V^4}}}\, ,
\eeqy 
where 
\beqy
\label{24}
V = \sqrt{v_\ell^2 + v_\phi^2 +g_{\rm mix}^2}\, .
\eeqy 

The speed of the transverse lattice phonon in Eq.~(\ref{17}) is unaffected by mixing and is approximately 
given by~\cite{cad92} 
\beqy
\label{25}
v_t \approx 0.4 \frac{\omega_p}{q_{\rm D}}  \approx 0.12 \left(\frac{Z}{50}\right)^{1/3} v_\ell
\, .
\eeqy

Note that due to entrainment effects, the expressions (\ref{20}), (\ref{22}) and (\ref{25}) for the velocities of the BA bosons 
and lattice phonons differ from those obtained considering either a neutron superfluid alone or a pure solid crust, respectively.
The self-consistent inclusion of entrainment is an important new element of this study. 

In the normal phase, any relative motion between the neutron liquid and the crust will be damped by collisions so in the hydrodynamic 
regime ions, electrons and neutrons will be essentially comoving. In this case, the Josephson's equation have to 
be replaced by the condition $\delta \pmb{v_n} = \delta \pmb{v_p}$. As a result, only one longitudinal mode corresponding to ordinary 
hydrodynamic sound persists and its velocity is given by 
\beqy\label{26}
c_s =\sqrt{\frac{K+4S/3}{\rho}}\, ,
\eeqy
where $K$ is the total bulk modulus of the crust. It is related to the bulk modulus $\tilde K$ of the electron-ion 
system by 
\beqy\label{27}
K=\widetilde K + 2 n_n L + n_n^2 \frac{\partial \mu_n}{\partial n_n}\, .
\eeqy
Since $S\ll K$ (see, e.g., sec. 7.1 of Ref.~\cite{lrr}), the sound velocity can be approximately written as
\beqy\label{28}
c_s \approx\sqrt{\frac{\partial P}{\partial \rho}} \approx
\sqrt{\frac{n_p}{n}~\frac{n_p}{m}~\frac{\partial \mu_e}{\partial n_e} +
\frac{n_n}{n}~\frac{n_n}{m}~\frac{\partial \mu_n}{\partial n_n}}\, .
\eeqy
The transverse mode velocity is given by 
\beqy\label{29}
v_t=\sqrt{\frac{S}{\rho}}\, .
\eeqy
While the existence of two weakly damped longitudinal modes is unique to the superfluid phase, entrainment is fairly insensitive 
to superfluidity provided the pairing gap $\Delta \ll \mu_n$~\cite{cch05b}, which is the case in most of the inner crust~\cite{bal07,cha10,for10}. 

%%%%%%%%%%%%%%%%%%%%%%%%%%%%%%%%%%%%%%%%%%%%%%%%%%%%%%%%%%%%%

%%%%%%%%%%%%%%%%%%%%%%%%%%%%%%%%%%%%%%%%%%%%%%%%%%%%%%%%%%%%%
\section{Microscopic model for the inner crust of a neutron star}  
\label{crust}
%%%%%%%%%%%%%%%%%%%%%%%%%%%%%%%%%%%%%%%%%%%%%%%%%%%%%%%%%%%%%

\begin{figure}
\includegraphics[scale=0.4]{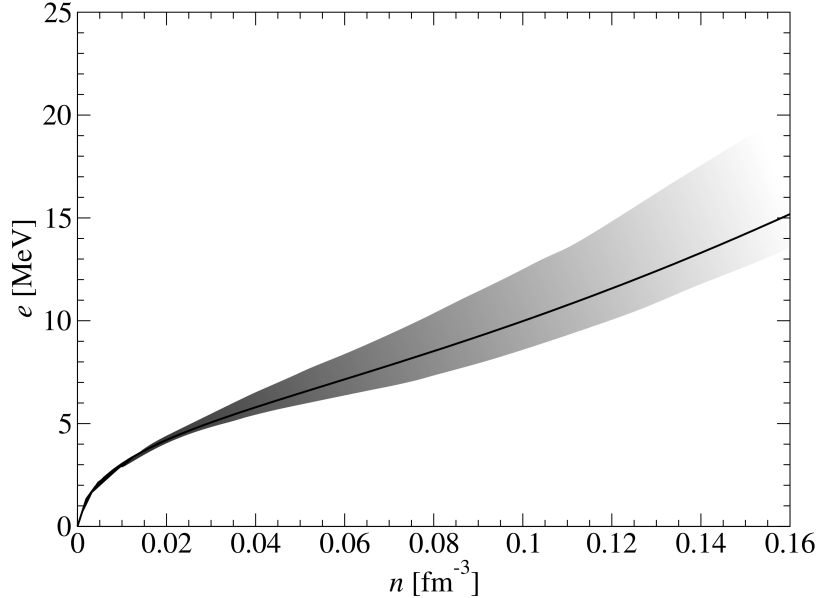}
\vskip -0.5cm
\caption{Energy per baryon in pure neutron matter as calculated by the Skyrme force BSk14~\cite{sg07} (solid line) and as obtained from 
next-to-next-to-next-to-leading order in chiral effective field theory~\cite{tew13} (shaded area).
}
\label{figNeuM}
\end{figure}

The evaluation of the velocities of the collective modes requires the knowledge of the susceptibilities defined by 
$\partial n_e/\partial \mu_e$ and $\partial n_n/\partial \mu_n$, and number densities $n_p$ and $n^{\rm b}_n$ for each 
given baryon density $n$. At densities above $\sim 10^6$ g cm$^{-3}$, electrons can be treated as an ideal relativistic 
Fermi gas so
\beqy\label{30}
\frac{\partial n_e}{\partial \mu_e}\approx \frac{3 n_e}{\mu_e} \, .
\eeqy
Electric charge neutrality requires $n_e=n_p$ so both $n_p$ and $\partial n_e/\partial \mu_e $ are uniquely determined 
by the composition of the inner crust (i.e., the variation of the electron density $n_e$ with $n$), 
taken from Ref.~\cite{onsi08}. 
The inner crust was assumed to be made of ``cold catalyzed matter'', i.e., matter in full thermodynamic equilibrium at 
zero temperature. Nuclei were supposed to be spherical, an assumption that is generally 
satisfied in all regions of the inner crust, except possibly near the crust-core interface where so-called nuclear ``pastas'' 
might exist (see, e.g., Sec. 3.3 in Ref.~\cite{lrr} for a brief review). The composition of the crust was obtained from a 
nonrelativistic Skyrme effective nuclear Hamiltonian solved using the fourth-order extended Thomas-Fermi method with proton 
quantum shell effects added via the Strutinsky-Integral theorem. Neutron quantum shell corrections, which were shown 
to be much smaller than proton quantum shell corrections~\cite{oya94,chamel2007}, were neglected. This so-called ETFSI method 
is a high-speed approximation to the self-consistent Skyrme-Hartree-Fock equations~\cite{sto07}. These calculations were carried out with the 
Skyrme force BSk14 underlying the HFB-14 atomic mass model~\cite{sg07}, which yields an excellent fit to essentially all experimental 
atomic mass data with a root mean square deviation of 0.73 MeV. At the same time, an optimal fit to charge radii was ensured.
Moreover, the incompressibility $K_v$ of symmetric nuclear matter at saturation was required to fall in the experimental range 
$240\pm10$~MeV~\cite{col04}. The symmetry energy $J$ and its slope $L$ play a crucial role for determining the structure of 
neutron-star crusts~\cite{grill2012}. The values predicted by the force BSk14, $J=30$~MeV, and $L=44$~MeV, respectively, are 
consistent with various constraints inferred from both experiments and astrophysical observations~\cite{lattimer2012}. For these 
reasons, the force BSk14 is expected to be well-suited for describing the nuclei in the inner crust of a neutron star. 
In addition, the BSk14 force was constrained to reproduce various properties of homogeneous nuclear matter as obtained from 
many-body calculations using realistic two- and three- nucleon interactions. In particular, the force BSk14 was fitted to the 
equation of state of neutron matter, as calculated by Friedman and Pandharipande~\cite{fp81} using realistic two- and three-body 
forces. Incidentally, this equation of state is in good agreement with more recent {\it ab initio} calculations~\cite{apr98,ger10,heb10,tew13} 
at densities relevant to the neutron-star crusts, as shown in Fig.\ref{figNeuM}. Therefore, the properties of the neutron liquid in the 
inner crust of a neutron star are well described by the Skyrme force BSk14. The crustal composition obtained in Ref.~\cite{onsi08} is 
summarized in Table~\ref{tab1}. 

%--------------------------------------------------------------------------------------------------------------------------------
%---------------------------------------------------------------------------------------------------------------------------------

\begin{table}
\begin{tabular}{|c|c|c|c|c|c|}
\hline
$n$ (fm$^{-3}$) & $\rho$ (g cm$^{-3}$) &$Z$ & $A_{\rm cell}$ & $A$ & $A^\star$  \\
\hline
0.0003 & $4.98\times10^{11}$ & 50 &   200 & 170 &   175  \\
0.001   & $1.66\times10^{12}$ & 50 &   460 & 179 &   383  \\
0.005   & $8.33\times10^{12}$ & 50 & 1140 & 198 &   975  \\
0.01     & $1.66\times10^{13}$ & 40 & 1215 & 170 & 1053  \\
0.02     & $3.32\times10^{13}$ & 40 & 1485 & 180 & 1389  \\
0.03     & $4.98\times10^{13}$ & 40 & 1590 & 173 & 1486  \\
0.04     & $6.66\times10^{13}$ & 40 & 1610 & 216 & 1462  \\
0.05     & $8.33\times10^{13}$ & 20 &   800 &   87 &   586  \\
0.06     & $1.00\times10^{14}$ & 20 &   780 &   85 &   461  \\
0.07     & $1.17\times10^{14}$ & 20 &   714 &   76 &   302  \\
0.08     & $1.33\times10^{14}$ & 20 &   665 &   65 &   247  \\
\hline
\end{tabular}
\caption{Ground-state composition  of the inner crust of a neutron star ($Z,A_{\rm cell},A$ as defined in Section~\ref{notations}), 
as obtained in Ref.~\cite{onsi08}, for various baryon densities $n$/mass densities $\rho$. The effective number of bound nucleons 
$A^\star$ was calculated including band structure effects in Ref.~\cite{cha12}. The density $n_n^{\rm b}$ of effectively bound 
neutrons can be obtained from Eq.~(\ref{2}). The density of conduction neutrons can be found from Eq.~(\ref{1}).} 
\label{tab1}
\end{table}

\begin{table}
\begin{tabular}{|c|c|c|c|c|c|c|c|c|c|}
\hline
$n$ (fm$^{-3}$) &  $g_{\rm mix}$ ($10^{-2}c$) & $v_\phi$ ($10^{-2}c$) & $v_\ell$ ($10^{-2}c$) & $c_s$ ($10^{-2}c$) &  $v_t$ ($10^{-2}c$) & $v_{-}$ ($10^{-2}c$) & $v_{+}$ ($10^{-2}c$)& $\lambda_-/\lambda_\ell$ & $\lambda_+/\lambda_\ell$ \\
\hline
0.0003 &  2.11 & 1.11(1.22)   & 5.13(5.21)  & 5.35 & 0.58(0.59) & 1.02 & 5.56 & 32.3 &  1.09 \\
0.001   & 2.60 & 1.34(2.56)  & 3.69(5.40)   & 4.46 & 0.42(0.61) & 1.08 & 4.58 & 9.12 &  1.28 \\
0.005   &  3.79 & 1.64(3.93)  & 2.60(5.76)  & 4.78 & 0.29(0.65) & 0.89 & 4.80 &  3.98 &  2.02 \\
0.01     & 4.34 & 1.77(4.49)  & 2.39(5.95)  & 5.18 & 0.25(0.62) & 0.81 & 5.20 &  3.63 &  2.40 \\
0.02     & 5.22 & 1.42(5.21)  & 2.26(6.28)  & 5.84 & 0.24(0.66) & 0.55 & 5.84 &  4.80 &  2.74 \\
0.03     & 5.95 & 1.62(5.97)  & 2.31(6.78)  & 6.55 & 0.24(0.71) & 0.57 & 6.56 &  4.59 &  3.01 \\
0.04     & 6.67 & 2.18(6.69)  & 2.44(6.36)  & 7.39 & 0.26(0.67) & 0.72 & 7.39 &  3.76 &  3.29 \\
0.05     &  6.73 & 4.21(7.69)  & 2.83(7.35) & 8.30 & 0.24(0.61) & 1.44 & 8.31 &  2.16 &  3.84 \\
0.06     &  6.73 & 5.86(8.65)  & 3.31(7.72) & 9.28 & 0.28(0.64) & 2.09 & 9.29 &  1.72 &  4.45 \\
0.07     &  6.20 & 7.76(9.66)  & 4.26(8.51) & 10.3 & 0.35(0.71) & 3.21 & 10.3 &  1.45 &  5.06 \\
0.08     &  6.34 & 8.98 (10.9) & 4.87(9.48) & 11.4 & 0.40(0.79) & 3.84 & 11.4 &  1.37 &  5.50 \\
\hline
\end{tabular}
\caption{Properties of collective modes in the inner crust of a neutron star. The velocities ($v_\phi$, $v_\ell$, $c_s$, $v_t$, $v_-$, 
and $v_+$) and the mixing parameter $g_{\rm mix}$ are defined in Section~\ref{dynamics}; $c$ is the speed of light. Values in 
parenthesis are obtained by neglecting entrainment. The ratios of the mean free path of the longitudinal modes to that of the 
unmixed longitudinal lattice phonon are shown in the last two columns.}
\label{tab2}
\end{table}

As discussed in detail in an accompanying paper~\cite{cha12}, neutron band-structure calculations are needed to determine 
$n^{\rm c}_n$. Here, we note that the key ingredient is the single-particle (s.p.) dispersion relation $\varepsilon_{\alpha\pmb{k}}$ 
($\alpha$ being the band index and $\pmb{k}$ the Bloch wave vector) given by the solution of the Schr\"odinger equation with the periodic 
mean-field potential obtained self-consistently from the ETFSI method. The superfluid density was then found from the equation
\beqy
\label{31}
n_n^{\rm c}=\frac{m}{24\pi^3\hbar^2}\sum_\alpha \int_{\rm F} |\pmb{\nabla}_{\pmb{k}} 
\varepsilon_{\alpha\pmb{k}}|{\rm d}{\cal S}^{(\alpha)}\, ,
\eeqy
where ${\rm d}{\cal S}^{(\alpha)}$ is an infinitesimal area element of the piecewise Fermi surface associated with the $\alpha$ band. 
As described in Ref.~\cite{cha12}, in most regions of the inner crust only a small fraction of dripped neutrons contributes 
to the superfluid density due to Bragg scattering so  $n_n^{\rm c} \ll n_n^{\rm f}$ or, equivalently $A^\star \approx A_{\rm cell}$. 
Note that unbound (bound) neutrons with density $n_n^{\rm f}$ (respectively, $n_n-n_n^{\rm f}$) are characterized by s.p. energies 
$\varepsilon_{\alpha\pmb{k}}$ lying above (respectively below) the largest value of the periodic mean-field potential. 
Results are summarized in Table~\ref{tab1}. 

The neutron chemical potential is determined by the neutron band structure from the equation
\beqy
\label{32}
n_n=\int_{-\infty}^{\mu_n} {\rm d}{\varepsilon} D(\varepsilon)\, ,
\eeqy
where $D(\varepsilon)$ is the density of neutron s.p. states defined by 
\beqy
\label{33}
D(\varepsilon)=\sum_\alpha \int \frac{{\rm d}^3\pmb{k}}{(2\pi)^3}\delta(\varepsilon-\varepsilon_{\alpha\pmb{k}})\, ,
\eeqy
where the $\pmb{k}$-space integration is taken over the first Brillouin zone. Differentiating Eq.~(\ref{32}) with respect to $\mu_n$ thus yields 
the neutron number susceptibility
\beqy
\label{34}
\frac{\partial n_n}{\partial \mu_n}=D(\mu_n) + \int_{-\infty}^{\mu_n} {\rm d}{\varepsilon} \frac{\partial D(\varepsilon)}{\partial \mu_n}\, .
\eeqy
Because nuclei in the inner crust are neutron saturated, the neutron susceptibility is essentially independent of the 
neutron bound states except possibly in a small region close to neutron drip. For the reasons explained in Ref.~\cite{cha09}, 
the density $D(\varepsilon)$ of neutron unbound states in a given region of the inner crust is well approximated by the density of 
s.p. states in uniform neutron matter for the corresponding density $n_n^{\rm f}$ of dripped neutrons. Using these approximations, 
the velocity of the BA mode in the inner crust can be expressed as
\beqy
\label{35}
v_\phi = \sqrt{\frac{n_n^{\rm c}}{n_n^{\rm f}}} v^{\rm f}_\phi\, ,
\eeqy
where $v^{\rm f}_\phi$ is the velocity of the BA mode in pure neutron matter at the density $n_n^{\rm f}$ associated with 
the crustal layer under consideration. This latter velocity is given by~\cite{leg66}
\beqy
\label{36}
v^{\rm f}_\phi=\frac{v_{\rm F}^2}{3}\left(1+F_0\right)\left(1+\frac{F_1}{3}\right)\, ,
\eeqy
where $v_{\rm F}$ is the Fermi velocity in pure neutron matter at the density $n_n^{\rm f}$ while $F_0$ and $F_1$ are the corresponding 
dimensionless Landau parameters, whose expressions for Skyrme interactions can be found in Ref.~\cite{gcp10}. We have evaluated 
$v^{\rm f}_\phi$ using the same Skyrme effective interaction BSk14 as that used to determine the equilibrium composition of the crust.

%--------------------------------------------------------------------------------------------------------------------------------
\begin{figure}[h]
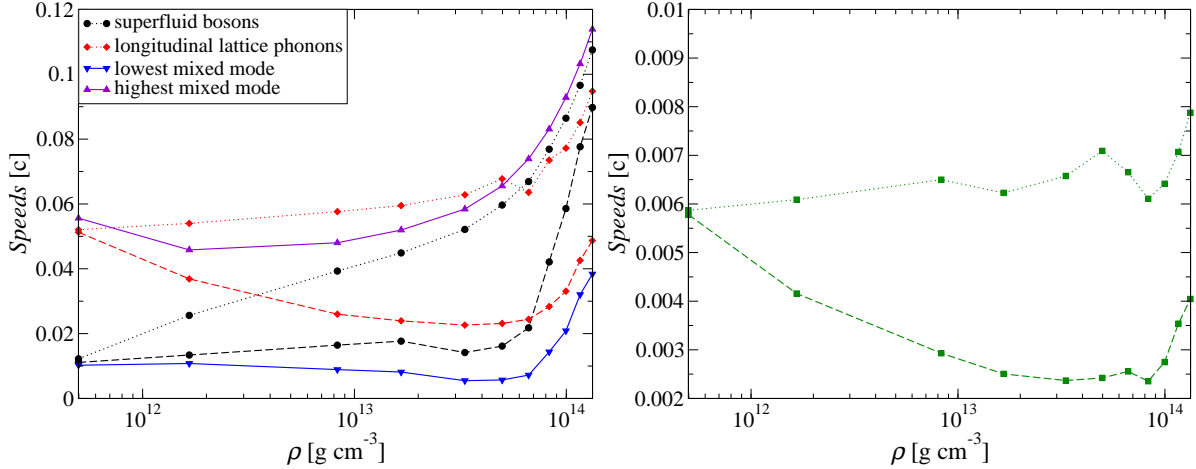

\includegraphics[scale=0.3]{long_speeds_r}
\includegraphics[scale=0.3]{trans_speeds_r}
\vskip -0.5cm
\caption{(Color online) Speeds (in units of the speed of light $c$) of the longitudinal (left panel) and transverse (right panel) collective excitations 
in the inner crust of a neutron star. Dotted curves show results with neither mixing nor entrainment, dashed curves include effects due 
to entrainment only and solid curves include in addition the effects due to mixing.
}
\label{fig1}
\end{figure}
%---------------------------------------------------------------------------------------------------------------------------------

The speeds of the collective modes in the inner crust  of a neutron star are shown in Fig.~\ref{fig1}, and listed in the Table ~\ref{tab2}. 
Entrainment modifies the spectrum: $v_\phi$, $v_\ell$, and $v_t$ are all significantly reduced (compare dotted and dashed curves), and mixing 
leads to a strong splitting between the longitudinal eigenmodes (note the difference between speeds of the lowest and highest eigenmodes). 
With increasing density, a strong suppression of the  plasma frequency due to entrainment leads to rapid decrease in the velocity of transverse
and longitudinal lattice phonon modes. Mixing between longitudinal modes leads to a high velocity eigenmode with velocity $v_+$ and a low velocity 
mode with velocity $v_-$. The  $v_-$ mode is predominantly the superfluid phonon (BA) mode near neutron drip and transforms to a mode with a 
large lattice component at the crust-core boundary. The mode with velocity $v_+$ is a pure lattice mode at neutron drip and transforms to being a 
mode which is predominantly a superfluid mode at the crust-core interface. 

With increasing temperatures, the neutron superfluidity may disappear in some regions of the crust. In these regions, the two longitudinal 
modes will merge and give rise to ordinary sound as discussed at the end of Sec.~\ref{dynamics}. Note, however, that the values for the 
speeds of collective excitations indicated in Table~\ref{tab2} are expected to remain essentially the same for temperatures $T\lesssim 10^{10}$~K.
Indeed, as shown in Ref.~\cite{onsi08}, thermal effects have a minor impact on the equilibrium composition of neutron-star crusts in this 
temperature range. However, the crust of a real neutron star may not necessarily be in full thermodynamic equilibrium, as discussed, e.g., 
in Sec. 3.4 of Ref.~\cite{lrr}. This could affect the spectrum of collective modes.

%%%%%%%%%%%%%%%%%%%%%%%%%%%%%%%%%%%%%%%%%%%%%%%%%%%%%%%%%%%%%
\section{Dissipation} 
\label{dissipation}
%%%%%%%%%%%%%%%%%%%%%%%%%%%%%%%%%%%%%%%%%%%%%%%%%%%%%%%%%%%%%
Lattice phonons couple strongly to electrons and easily excite electron-hole pairs in the dense electron gas. This Landau damping 
of lattice phonons has been studied in Ref.~\cite{chu07} and an approximate result of the lattice phonon mean free path was obtained. 
The mean free path of a thermal phonon that contributes to thermal conductivity was found to be  
\beqy
\label{37}
\lambda_{\text{lph}}=\frac{6\pi}{Ze^2~\gamma~\bar{v}}~\frac{1}{q_{\rm D}}~\frac{F(T_p/T)}{\Lambda_{\rm ph-e}}\simeq 72.5~\left(\frac{40}{Z}\right)^{2/3}~\left(\frac{F(T_p/T)}{\bar{v}~\Lambda_{\rm ph-e}}\right)~r_\mathrm{cell}\, ,
\eeqy
where 
\beqy 
\label{38}
F(T_p/T)=0.014 + \frac{0.03}{\exp{(T_p/(5T))}+1} \, ,  
\eeqy
\beqy\label{39}
\Lambda_{\rm ph-e}= \ln{\left(\frac{2}{\gamma}\right)}-\frac{1}{2}\left(1-\frac{\gamma^2}{4}\right)\quad {\rm and} 
\quad \gamma=q_{\rm D}/k_{\rm Fe}\, ,
\eeqy
and $\bar{v}$ is average velocity of the lattice phonon. Note that for simplicity we have neglected corrections due to the Debye-Waller 
factor and the nuclear form factor to the Coulomb logarithm $\Lambda_{\rm ph-e}$. Such corrections tend to increase the mean free path and 
Eq.~(\ref{37}) therefore must be viewed as a lower limit. 

Our interest here is to investigate the mean free path of the superfluid phonon mode in the inner crust. In Ref.~\cite{agu09} it was shown 
that phonon-phonon and phonon-impurity scattering were negligible compared to the dissipation that arose due to mixing with the lattice 
phonon. The superfluid phonon mean free path, without the inclusion of entrainment effects, was found to be much larger than that of the 
lattice phonons because mixing due to the density interaction was weak. In the following, we include effects due to entrainment, which is 
now known to be the dominant contribution to the mixing parameter $g_{\rm mix}$, and we show that the mean free path of the superfluid mode 
is greatly reduced due to strong mixing. Incorporating this into the dispersion relation in Eq.~(\ref{18}) we obtain
\beqy
\label{40}
(\omega^2-v_\phi^2~q^2)(\omega^2-2{\rm i}\Gamma_\ell~ \omega-v_\ell^2~q^2)=g^2_{\rm mix} ~\omega^2~q^2\,, 
\eeqy
where $\Gamma_\ell = v_\ell/\lambda_\ell$  and $\lambda_\ell$ is the mean free path of the lattice phonon in the limit of weak 
damping ($\Gamma_\ell \ll v_\ell q$). In general, $\lambda_\ell \ne \lambda_{\text{lph}}$ as the latter is an average mean free path more
closely related to the mean free path of the transverse thermal phonon. Nonetheless it provides an order of magnitude estimate. 

Mode mixing induces an indirect coupling between the superfluid BA bosons and electrons. Because the longitudinal modes contain an 
admixture of superfluid and lattice phonons, the damping of lattice vibrations due to electron-hole excitations naturally leads to 
a finite damping of both modes. 
In a small region of the crust in the vicinity of the neutron-drip transition where $g^2_{\rm mix} \ll v^2_\ell  - v^2_\phi $ the modes 
are not strongly mixed, and using the fact that $v_\ell \gg v_\phi$, we can obtain from Eq.~(\ref{40}) the analytic relation 
\beqy
\label{41}
\lambda_\phi \approx \frac{v_\ell^3}{g_{\rm mix}^2~v_\phi}~\lambda_\ell =  \left(\frac{v_\ell}{v_\phi}\right)^3~\frac{n^{\rm c}_n(n_p+n^{\rm b}_n)}{(n^{\rm b}_n)^2} ~\lambda_\ell \gg \lambda_\ell \,,
\eeqy 
between the mean free paths of the superfluid and lattice modes. In other regions mixing is strong and damping associated with each eigenmodes 
is found by solving Eq.~(\ref{40}). Although an analytic solution exists, it is cumbersome to write down explicitly. We present numerical
values for the ratio of the mean free paths $\lambda_+/\lambda_\ell$ and $\lambda_-/\lambda_\ell$ where $\lambda_\pm$ are the mean free paths 
of the eigenmodes in the last two columns of  Table~\ref{tab2}. It is meaningful to calculate these ratios without specifying $\lambda_\ell$ 
because it is independent of $\lambda_\ell$ in the weak damping limit. From the table we see that the mean free path of the mode with a large 
superfluid component is large near neutron drip, but decreases rapidly due to mixing when $v_\ell \simeq v_\phi$. In the bulk of the inner 
crust both modes have comparable mean free paths and this behavior qualitatively differs from that observed in Ref.~\cite{agu09} where 
entrainment was neglected and mixing was found to be weak except in a narrow region close to resonance.      

%%%%%%%%%%%%%%%%%%%%%%%%%%%%%%%%%%%%%%%%%%%%%%%%%%%%%%%%%%%%%
\section{Implications} 
\label{implications}
%%%%%%%%%%%%%%%%%%%%%%%%%%%%%%%%%%%%%%%%%%%%%%%%%%%%%%%%%%%%%

X-ray observations of accreting neutron stars in low-mass x-ray binaries have recently proved to be very useful 
for probing neutron-star interiors.
The accretion of matter onto the surface of the neutron star triggers thermonuclear fusion reactions. 
Under certain circumstances, these reactions can become explosive, giving rise to x-ray bursts and superbursts~\cite{stro06}.
The ignition conditions of these thermonuclear flashes depend sensitively on the thermal properties of the crust.
Valuable information on neutron star crusts can also be obtained from the thermal x-ray emission in quiescence
following a long outburst of accretion during which the crust has been driven out of its thermal equilibrium with the core~\cite{rut02}.
The thermal relaxation between the accreting and quiescent stages has been monitored for the four quasipersistent 
soft x-ray transients KS 1731$-$260~\cite{ca10}, MXB 1659$-$29~\cite{ca08}, XTE J1701$-$462~\cite{fr11}, and 
EXO 0748$-$676~\cite{de11}. Numerical simulations of these phenomena have shown that the cooling is very sensitive to the 
properties of the neutron-star crust~\cite{rut02,sht07,bc09}.
In particular, the thermal relaxation time of the crust is approximately given by~\cite{la94,gne01} 
\beqy\label{42}
\tau \sim (\Delta R)^2 \left(1-\frac{2 G M}{R c^2}\right)^{-3/2} \frac{C_V}{\kappa}
\eeqy
where $\Delta R$ is the crust thickness, $R$ is the radius, and $M$ is gravitational mass of the neutron star, 
while $C_V$ and $\kappa$ are the average heat capacity and thermal conductivity in the density range between 
$\sim 0.1 \bar n_{\rm cc}$ and  $\bar n_{\rm cc}$ where $\bar n_{\rm cc}=0.08$ fm$^{-3}$ is the crust-core transition 
density. The thermal relaxation of hot newly born neutron stars could also shed light on the crust properties. However, 
such very young neutron stars have not been observed yet, being presumably obscured by their expanding supernova envelope. 

%--------------------------------------------------------------------------------------------------------------------------------
\begin{figure}[ht]
\includegraphics [scale=0.40]{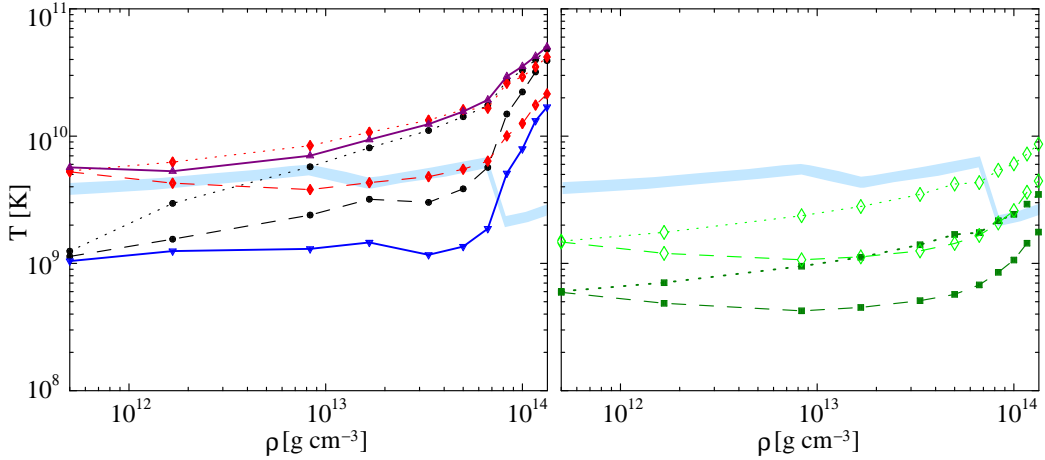}  
\caption{(Color online) Left panel: Debye temperatures $\Theta_{D}$ of the longitudinal collective excitations using the 
same notations as in Fig.~\ref{fig1}.
Dotted curves show values with neither mixing nor entrainment, dashed curves include effects due to
entrainment only and solid curves include in addition the effects due to mixing.
Right panel: Debye temperature $\Theta_{D}$ of the transverse collective modes (lines with filled squares)
and ion plasma temperature $T_p = (\hbar/k_\mathrm{B}) \omega_p$ (lines with diamonds). Dotted (dashed) curves show values
without (with) entrainment. In both panels, the light blue band delimits the range of $T$ below which nuclei crystalize
(using $\Gamma_c = 180$ to $220$).
}
\label{Fig:T_D}
\end{figure}
%--------------------------------------------------------------------------------------------------------------------------------
\begin{figure}[ht]
\includegraphics [scale=0.32]{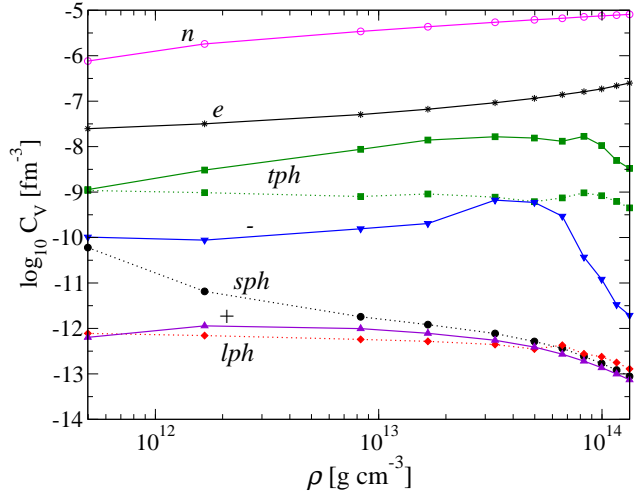}
\caption{(Color online) Heat capacity of electrons ($e$), transverse lattice phonons ($tph$) and longitudinal excitations ($-$ and $+$) 
in the inner crust of neutron stars at $T=10^7$~K, with mixing and entrainment effects (solid lines) and without (dotted lines). In the 
absence of mixing, the longitudinal modes are the Bogoliubov-Anderson superfluid phonons ($sph$) and the longitudinal lattice phonons ($lph$). 
For comparison, is also shown the normal neutron contribution ($n$), but it is strongly suppressed by superfluidity except
in the shallowest and densest parts of the inner crust where the neutron $^1$S$_0$ pairing gap becomes vanishingly small.}
\label{fig2a}
\end{figure}
%--------------------------------------------------------------------------------------------------------------------------------
\begin{figure}[ht]
\includegraphics [scale=0.32]{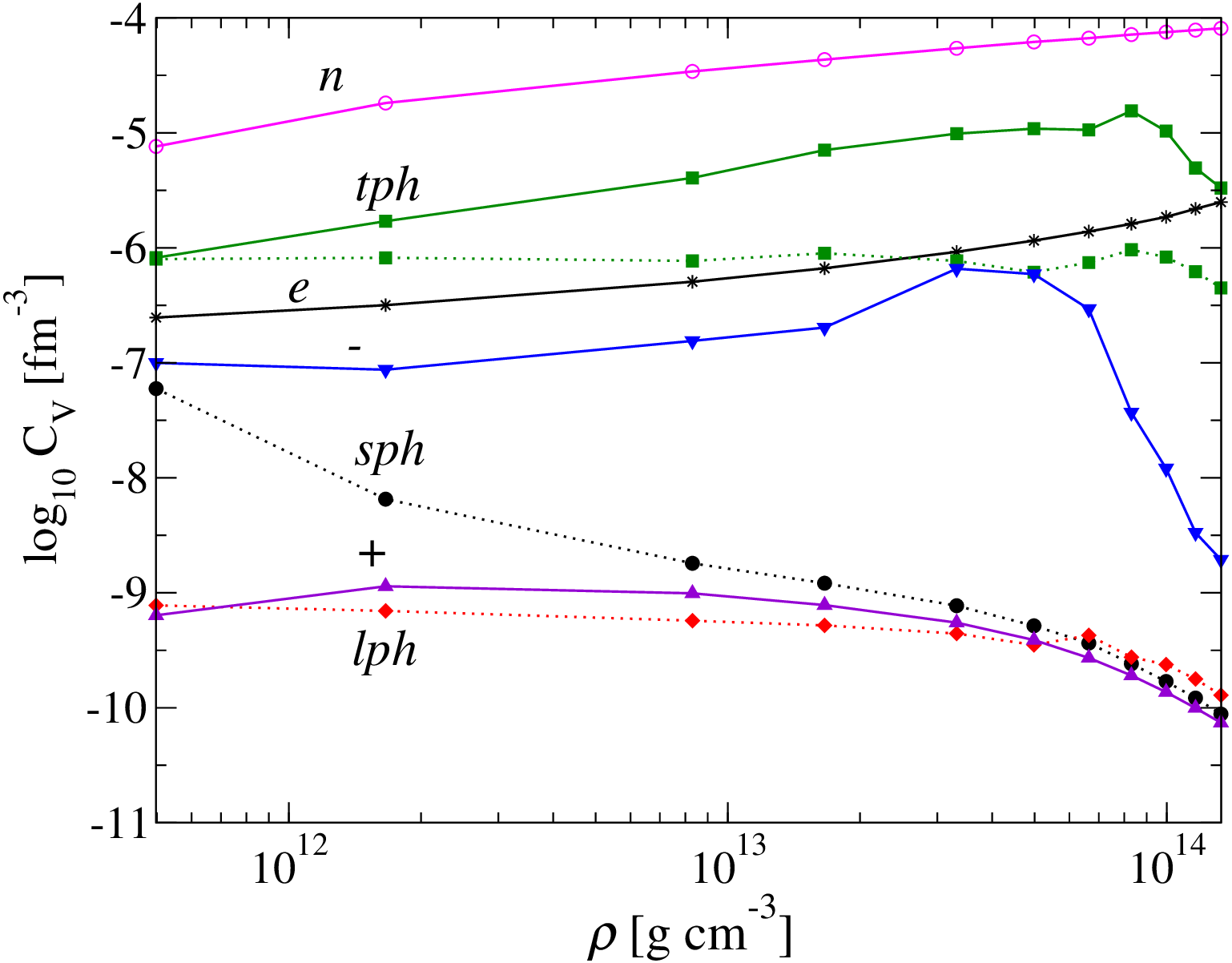}
\caption{(Color online) Same as Fig.~\ref{fig2a} for $T=10^8$~K.}
\label{fig2b}
\end{figure}
%--------------------------------------------------------------------------------------------------------------------------------
\begin{figure}[ht]
\includegraphics [scale=0.32]{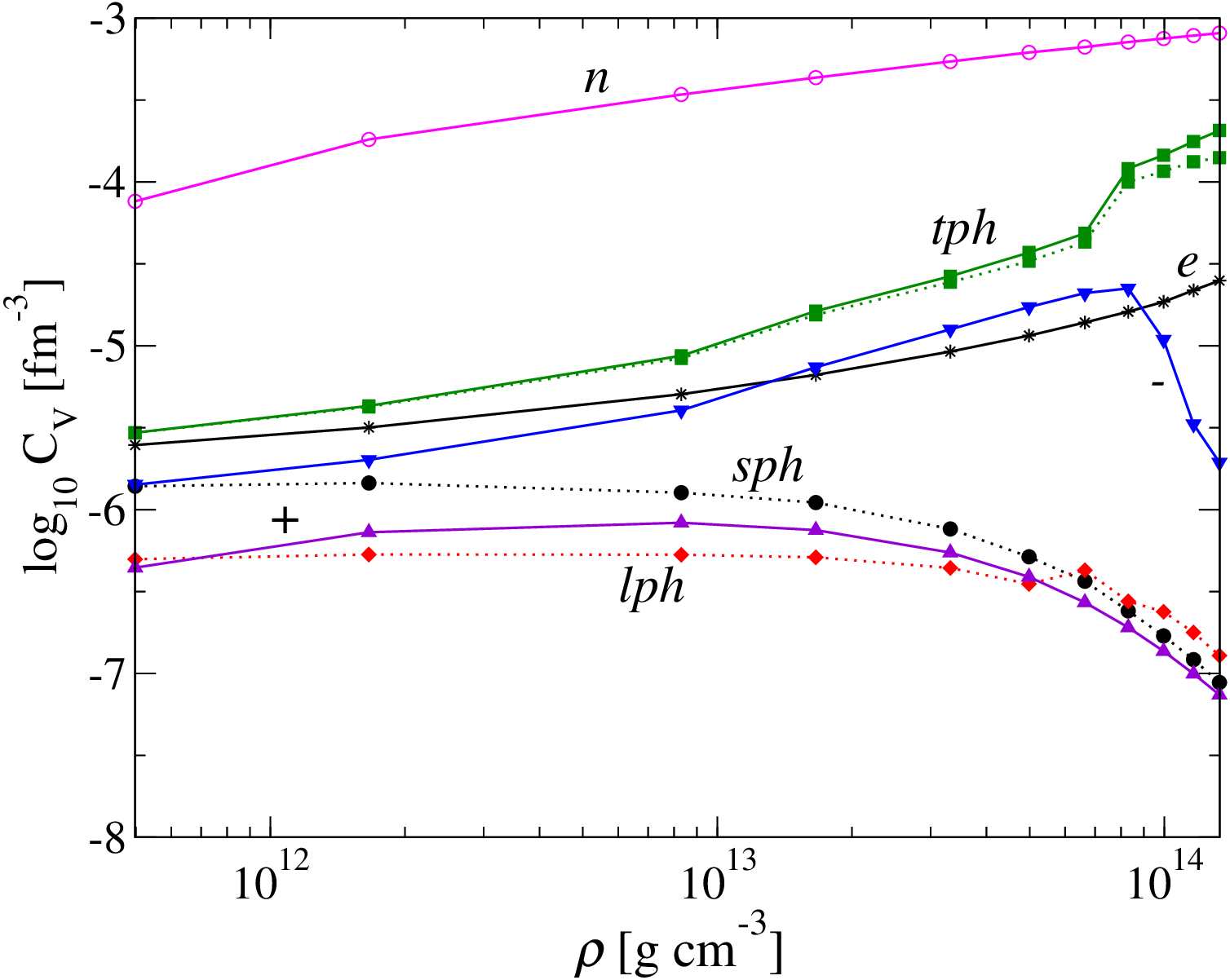}
\caption{(Color online) Same as Fig.~\ref{fig2a} for $T=10^9$~K..}
\label{fig2c}
\end{figure}
%--------------------------------------------------------------------------------------------------------------------------------

The inner crust heat capacity is the sum of contributions from the quasiparticle excitations of the electron gas and
neutron liquid, and from the collective excitations described above. In what follows, we describe these contributions to the 
volumetric crustal heat capacity. 
Treating electrons as a relativistic Fermi gas, their heat capacity is simply given by ($k_{\rm B} T\ll \mu_e$)
\beqy
\label{43}
C_V^e = \frac{1}{3} \frac{\mu_e^2}{(\hbar c)^3} k_{\rm B} T  \, .
\eeqy
The heat capacity of non-superfluid degenerate neutrons (for $k_{\rm B} T\ll \mu_n$) is similarly given by  
\beqy
C_V^n=\frac{1}{3} \pi^2 D(\mu_n) k_{\rm B} T\, ,
\label{44}
\eeqy 
where $D(\mu_n)$ is well approximated by the density of states in uniform neutron matter at the density $n_n^{\rm f}$~\cite{cha09}. 
This neutron contribution is enormous and will always dominate 
in the layers where neutrons are normal. Once superfluidity sets in, however, $C_V^n$ is strongly suppresssed and becomes 
negligible when the temperature is much lower than the critical temperature $T_c^n$~\cite{cha10,for10}. Given the density 
dependence of the neutron $^1$S$_0$ gap there are only two regions, just above the neutron drip point and possibly in the 
deepest part of the crust, where $C_V^n$ is relevant (see, e.g., Ref.~\cite{pr12}).

The heat capacity associated with a collective excitation having a dispersion relation of the form $\omega = v q$ is given by 
\beqy
\label{45}
C_V^{\rm coll} = \frac{3 n_p}{Z x_D^3}\int_0^{x_D}{\rm d}x \frac{x^4 e^x}{(e^x-1)^2}  \, ,
\eeqy
with $x_D = \Theta_D/T$, $\Theta_D = (\hbar/k_{\rm B}) \, q_D v$ being the Debye temperature of the collective mode.
At low temperatures, $T \ll \Theta_D$ such that $x_D\gg 1$, one has the standard Debye result
\beqy\label{46}
C_V^{\rm coll} \simeq \frac{2\pi^2}{15} \left(\frac{k_{\rm B}T}{\hbar v}\right)^3 =
n_\mathrm{I}  \, \frac{4\pi^4}{5} \left(\frac{k_{\rm B}T}{\hbar q_D v}\right)^3
\eeqy
while at high temperatures, $T \gg \Theta_D$ when $x_D\ll 1$, one obtains the classical result $C_V^{\rm coll}  = n_\mathrm{I}$.
At low-enough temperatures ($T\ll \Theta_D$ and $T\ll T_c^n$) the heat capacity of the crust is, hence, approximately given by
\beqy
\label{47}
C_V \simeq
\frac{1}{3} \frac{\mu_e^2}{(\hbar c)^3}k_{\rm B} T + \frac{2\pi^2}{15} \left(\frac{k_{\rm B} T}{\hbar \bar v}\right)^3  \, ,
\eeqy
with 
\beqy\label{48}
\frac{1}{\bar v^3} = \frac{2}{v_t^3}+ \frac{1}{v_{-}^3} +\frac{1}{v_{+}^3} \, .
\eeqy

We plot in Fig.~\ref{Fig:T_D} the Debye temperatures of the four collective modes and
the various contributions to $C_V$ are displayed in Figs.~\ref{fig2a},\ref{fig2b}, and \ref{fig2c} for three typical temperatures
of astrophysical interest. While at $T = 10^9$ K the $v_-$ mixed mode and the two degenerate transverse modes are in the classical
regime, all modes are well into the quantum regime at $T=10^8$~K and $T=10^7$~K. This suggests that entrainment 
and mixing will not affect the thermal relaxation of newly-born isolated neutron stars but could be important 
for accreting neutron stars.

Figures~\ref{fig2a}, \ref{fig2b}, and\ref{fig2c} show that the heat capacity of non-superfluid neutrons would largely dominate over all collective 
modes, but becomes insignificant once neutron superfluidity sets in, i.e., in most of the inner crust. Overall, the 
transverse lattice mode contribution $C_V^t$ to the heat capacity, dominates at $T=10^9$~K and $10^8$~K, while electrons 
dominate at $10^7$~K due to the linear temperature dependence of $C_V^e$ compared to the $T^3$ dependence for $C_V^t$.
$C_V^t$ in not affected by entrainment at  $T=10^9$~K, since the transverse modes are in the classical regime,
but at $T=10^8$~K and $10^7$~K it is increased by almost one order of magnitude in most of the crust. Notice that at 
$T=10^8$~K without entrainment $C_V^t$ would be comparable to $C_V^e$ while it clearly dominates once entrainment
is taken into account. Moreover, the heat capacity of the longitudinal mode is increased by several orders of magnitude 
by entrainment, and mixing. In particular, the contribution of the lowest mixed mode becomes even comparable with $C_V^e$ at 
high temperatures.

\begin{figure}
\includegraphics[scale=0.4]{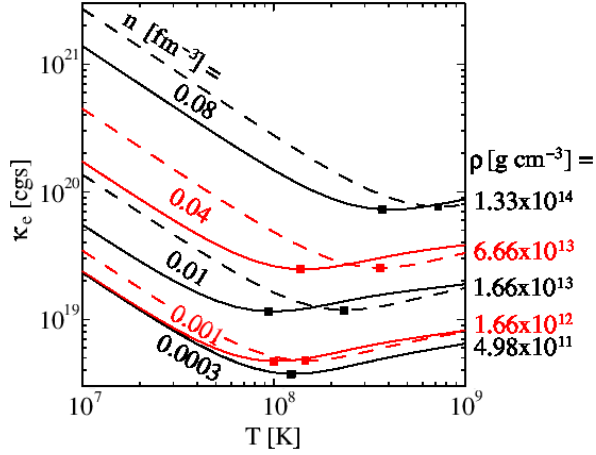}
%\vskip -0.5cm
\caption{Electron thermal conductivity $\kappa_e$ in the neutron-star crust, with (solid) and without (dashed) entrainment effects included,
at five densities $n =$ 0.0003, 0.001, 0.01, 0.02, and 0.08 fm$^{-3}$ as labeled on the curves.
The minimum of $\kappa_e$ occurs at $T \sim 0.1 T_p$, marked by a square on the corresponding curve; 
$\kappa_e \propto T^{-1}$ in the quantum regime, $T\ll T_p$, while it only weakly increases with $T$ in 
the classical regime at higher temperatures.
}
\label{fig:elec_conduct}
\end{figure}

Because entrainment modifies the spectrum of collective excitations, it also affects the heat transport in the crust. 
The thermal conductivity is generally governed by electrons. Changes of phonon velocities alter the electron-phonon process
hence also the electron thermal conductivity. The conductivity is mainly limited by the Umklapp process, in which an electron
simultaneously Bragg scatters off the lattice, and emits a transverse phonon~\cite{flo76,rai82}.
Since the scattering rate scales as $v^{-3}$, where $v$ is the phonon velocity, and $v_\ell \gg v_t$, processes involving longitudinal 
phonons are typically negligible. This observation also permits us to reliably estimate the changes in the electron mean free path due 
to entrainment. First, we note that the electron-phonon scattering rate depends on the electron Fermi momentum $k_{\rm Fe}$, the ion 
plasma frequency $\omega_p$ and $v_t$ (see Ref.~\cite{pr12} for a discussion). Since $v_t \propto \omega_p/q_{\rm D}$, it follows that 
effects due to entrainment on the scattering are entirely incorporated through its effects on $\omega_p$. It therefore suffices to 
employ an existing fitting formula developed in an earlier work but with a suitably reduced value of $\omega_p$ due to entrainment. 
In Fig.~\ref{fig:elec_conduct} we plot the electron thermal conductivity with and without entrainment effects included. As anticipated, 
the conductivity decreases with entrainment simply reflecting the fact that it is easier to excite lower velocity transverse phonon modes.

%--------------------------------------------------------------------------------------------------------------------------------
%---------------------------------------------------------------------------------------------------------------------------------

%--------------------------------------------------------------------------------------------------------------------------------
\begin{figure}
\includegraphics[scale=0.4]{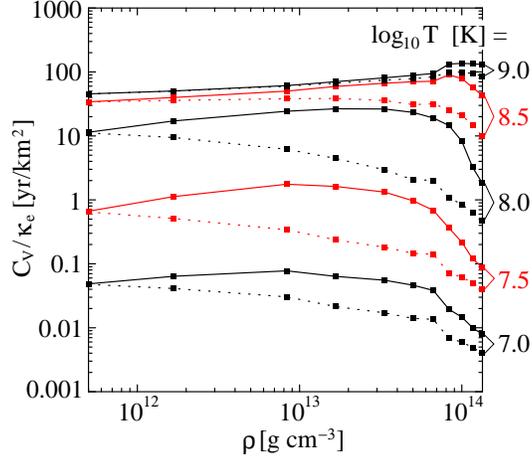}
%\vskip -0.5cm
\caption{Ratio of the heat capacity $C_V$ to the electron thermal conductivity $\kappa_e$ at five temperatures, as labeled.
$C_V$ includes the ion as well as the electron contributions, the neutron part is neglected. 
Continuous lines show values when entrainment, through its modification of $T_p$, 
is taken into account, while in the values for dotted lines it is neglected.
}
\label{fig:thermal_time}
\end{figure}
%---------------------------------------------------------------------------------------------------------------------------------

Having described the impact of entrainment on reducing the
electron thermal conductivity and increasing the lattice specific
heat, we now discuss their combined effect on the thermal
time scale, Eq. (\ref{42}). We plot $C_V/\kappa$ in Fig.~\ref{fig:thermal_time} for five different
temperatures. For $T = 10^9$~K, the impact of entrainment is
negligible since $T$ is comparable or larger than $\Theta_D$ of the
transverse modes, as already pointed out previously. As the
temperature is decreased, entrainment leads to a significant
enhancement in $C_V/\kappa$, hence also in $\tau$ : at $\rho = 10^{13}$~ g~cm$^{−3}$ and
for $T = 10^8$~K, $\tau$ can be increased by more than one order of
magnitude. For $T = 10^7$~K, the lowest temperature considered
here, the effect of entrainment is smaller, being moderated by
the dominance of the electron contribution to $C_V$.

Although electrons dominate heat conduction under normal conditions, phonons can contribute either at high temperature when 
$C^{\rm coll}_{V}\ge C^{e}_V$ or when large magnetic fields suppress electron conduction transverse to the field \cite{chu07,agu09}. 
In the inner crust, the lattice and superfluid phonons contributions were estimated in Ref.~\cite{chu07} and Ref.~\cite{agu09}, respectively. 
From kinetic theory and in the case where phonon conduction is diffusive (rather than convective), the thermal conductivity is given by 
\beqy\label{49}
\kappa^{\rm coll} = \frac{1}{3}~C^{\rm coll}_V~v~\lambda \,,
\eeqy 
where $C^{\rm coll}_V$ is the heat capacity, $v$ is the velocity, and $\lambda$ is the mean free path of each collective mode.   
Entrainment alters the thermal conductivity through these three factors. The larger specific heat associated with lower velocity transverse 
modes implies that their contribution to the heat conduction is proportionately enhanced. In addition, since $\lambda \propto 1/v$, 
the smaller $v_t$ acts to further increase the conductivity, and the combined effect is to increase the earlier estimate of 
Ref.~\cite{chu07} by the factor $(A^\star/A)^{3/2}$. 

The effects on the superfluid phonon contribution is more complex because mixing is strong throughout the inner crust except in the vicinity 
of neutron drip. It is only meaningful to discuss heat diffusion due to eigenmodes, and, in general, there are two competing effects due 
entrainment. At first entrainment lowers the velocity of the mode with a larger superfluid component and increases its heat capacity, but 
with increasing density this increase is overcome by strong mixing which dramatically reduces the mean free path. Since $\lambda_+$ and 
$\lambda_-$ are of the same magnitude as $\lambda_\ell$, and because $v_+ \gg v_t $ and $v_-\gg v_t$ their contribution to heat transport 
is typically negligible. This new result implies that superfluid modes may play a smaller role in heat transport in magnetars than 
anticipated in Ref.~\cite{agu09},  and it is likely that the enhanced heat conduction due to the transverse mode will dominate in much 
of the inner crust.

%%%%%%%%%%%%%%%%%%%%%%%%%%%%%%%%%%%%%%%%%%%%%%%%%%%%%%%%%%%%%
\section{Conclusions}
\label{conclusions} 
%%%%%%%%%%%%%%%%%%%%%%%%%%%%%%%%%%%%%%%%%%%%%%%%%%%%%%%%%%%%%

A large fraction of dripped neutrons in the inner crust of a neutron star are entrained by nuclei and move with them, due to coherent 
(Bragg) scattering of neutrons by the crystal lattice~\cite{cha05,cha12}. This nondissipative entrainment induces a strong coupling 
between the superfluid and lattice dynamics and is shown to affect the spectrum of low-energy collective excitations of the inner crust. 
Superfluid and longitudinal lattice phonons are found to be very strongly mixed, and the speed of transverse lattice modes is greatly 
reduced thus leading to a significant enhancement of the crustal specific heat at temperatures above $\sim 10^8$~K. This, combined with 
entrainment induced reduction in the electron mean free path, entails an increase of the heat diffusion time in the crust, especially for 
temperatures in the range $10^7-10^8$~K encountered in quasi-persistent soft x-ray transients. This warrants the need to take into account 
entrainment effects in the interpretation of the observed thermal relaxation in these accreting neutron stars.

Shear modes in neutron-star crusts with velocity in the range $v_t \simeq 10^{-3}-10^{-2}$ c have been proposed to play a role in the 
interpretation of quasiperiodic oscillations (QPOs) observed in giant flares from SGRs~\cite{str06}. The fundamental 
frequency of the global shear mode is given by $\Omega_0 \simeq \bar{v}_t  /2 \pi R$, $R$ being the neutron-star radius and $\bar{v}_t$ an 
appropriate average of the shear velocity in the inner crust, where the mode energy mainly resides~\cite{pir05}. Since entrainment lowers 
$v_t$ by a factor of about $2-3$ in most of the inner crust, our results suggest that $\Omega_0$  is too small to account for the observed 
QPO frequencies in the giant flares~\cite{ste09}. It is also likely that the existence of the low-velocity longitudinal eigenmode in the 
coupled superfluid-solid inner crust may be relevant to interpret global oscillation modes. 

However, there are several issues that deserve further attention before one can draw quantitative conclusions from our study. The possible 
presence of nuclear ``pastas'' in the deep regions of the inner crust, which has been neglected here, would reduce the effects of 
Bragg scattering ~\cite{cch05} and change the temperature dependence of the specific heat at low temperatures~\cite{dig11} due to the low 
dimensionality of these configurations. Besides, the composition and the properties of neutron-star crusts may differ from those of 
cold-catalyzed matter that we have considered in this work. We anticipate that quantum and thermal fluctuations of nuclei about their 
equilibrium positions, crystal defects, impurities, and, more generally, any source of disorder would presumably reduce the number of 
entrained neutrons. Quantitative estimates of all these effects is beyond the scope of this work.

%%%%%%%%%%%%%%%%%%%%%%%%%%%%%%%%%%%%%%%%%%%%%%%%%%%%%%%%%%%%%
\begin{acknowledgments}
This work was financially supported by FNRS (Belgium) and CompStar, a Research Networking Programme of the European Science Foundation. 
N.C. thanks the Institute for Nuclear Theory at the University of Washington for its hospitality and the Department of Energy for 
partial support. 
The work of S.R. was supported by DOE Grant No. \#DE-FG02-00ER41132 and by the 
Topical Collaboration to study {\it neutrinos and nucleosynthesis in hot dense matter}. 
D.P's work was supported by grants from Conacyt (Grant No. CB-2009/132400) and UNAM-DGAPA (Grant No. PAPIIT IN113211).
\end{acknowledgments}
%%%%%%%%%%%%%%%%%%%%%%%%%%%%%%%%%%%%%%%%%%%%%%%%%%%%%%%%%%%%%

%%%%%%%%%%%%%%%%%%%%%%%%%%%%%%%%%%%%%%%%%%%%%%%%%%%%%%%%%%%%%

%%%%%%%%%%%%%%%%%%%%%%%%%%%%%%%%%%%%%%%%%%%%%%%%%%%%%%%%%%%%%

%%%%%%%%%%%%%%%%%%%%%%%%%%%%%%%%%%%%%%%%%%%%%%%%%%%%%%%%%%%%%
%%%%%%%%%%%%%%%%%%%%%%%%%%%%%%%%%%%%%%%%%%%%%%%%%%%%%%%%%%%%%
\end{document}